\documentclass[a4paper]{jpconf}
\usepackage{graphicx}
\usepackage{amssymb}
\usepackage{amsmath}
\begin{document}

\title{Formation of plasmoid chains in fusion relevant plasmas}

\author{L Comisso$^{1}$, D Grasso$^{1}$ and F L Waelbroeck$^2$}

\address{{$^1$ Dipartimento Energia, Politecnico di Torino, Corso Duca degli Abruzzi 24, 10129, Torino, and Istituto dei Sistemi Complessi - CNR, Via dei Taurini 19, 00185, Roma, Italy}\\
{$^2$ Institute for Fusion Studies, The University of Texas at Austin, Austin, TX 78712-1060, USA}}

\ead{luca.comisso@polito.it}

\begin{abstract}
The formation of plasmoid chains is explored for the first time within the context of the Taylor problem, in which magnetic reconnection is driven by a small amplitude boundary perturbation in a tearing-stable slab plasma equilibrium. Numerical simulations of a magnetohydrodynamical model of the plasma show that for very small plasma resistivity and viscosity, the linear inertial phase is followed by a nonlinear Sweet-Parker evolution, which gives way to a faster reconnection regime characterized by a chain of plasmoids instead of a slower Rutherford phase.
\end{abstract}

\section{Introduction}
Magnetic reconnection is a fundamental plasma process that changes the topology of the magnetic field lines and has profound consequences in a wide variety of different phenomena in laboratory, space and astrophysical plasmas \cite{Yamada_2010}. It can be spontaneous and/or forced. In the latter case, one of the most important paradigms is the so-called Taylor problem, which consists in the study of the response of a tearing-stable slab plasma equilibrium to a small amplitude boundary perturbation that drives magnetic reconnection. This problem was investigated for the first time in a seminal work by Hahm and Kulsrud \cite{HK_1985}. Adopting a resistive magnetohydrodynamic (MHD) model of the plasma, they showed that the reconnection process evolves through five dynamical phases, the first four dominated by linear layer physics, and the fifth one characterized by a nonlinear Rutherford phase \cite{Rutherford_1973}.

The analysis by Hahm and Kulsrud was reconsidered some years later by Wang and Bhattacharjee, who showed that under certain conditions the reconnection process passes into the nonlinear regime according to a Sweet-Parker evolution which gives way, on the long time scale of resistive diffusion, to the Rutherford regime \cite{WB_1992}. Subsequently, an expression for the threshold perturbation amplitude required to trigger the Sweet-Parker phase in a visco-resistive plasma was derived in a careful work by Fitzpatrick \cite{Fitz_2003}.

The Taylor problem is of great interest within the fusion community, mainly because it represents a convenient model to describe magnetic reconnection processes due to resonant magnetic perturbations. 
However, despite the numerous works related to this problem (see, e.g., [6-14]), in the framework of resistive and visco-resistive MHD the nonlinear possible scenarios of the Taylor paradigm have not changed from those outlined in Refs. \cite{HK_1985,WB_1992,Fitz_2003}.

In this paper we show a different possible scenario, in which the Sweet-Parker phase identified by Wang and Bhattacharjee does not lead to a slower Rutherford evolution, but to a faster regime characterized by the formation of plasmoid chains. Actually, it is known from quite a long time that elongated current sheets can become unstable to the formation of plasmoids (secondary islands) [15-18]. 
Nevertheless, the eventuality that the current sheet may be unstable has never been taken into account whithin the context of the Taylor problem.

In the last years there has been a growing interest to investigate the plasmoid instability of thin current sheets due to the fact that the plasmoid formation facilitates faster reconnection [19-30]. 
Here we show that the development of plasmoids is possible also whithin the Taylor paradigm and that strongly affects the reconnection rate. Interestingly, a recent work by Dewar and coworkers \cite{Dewar_2013} have found the existence of MHD equilibria with plasmoids as solutions of the Taylor geometry. However, this work was concerned only with finding static equilibrium solutions.

\section{Basic equations and geometry of the viscous Taylor problem}  \label{sec2}

We consider a visco-resistive MHD description of the plasma. Within this description the equation of motion of the plasma takes the form
\begin{equation}
\begin{array}{*{20}{c}}
\, \\
\, 
\end{array} 
\rho \left( {\partial_t + {\bf{v}} \cdot \nabla } \right){\bf{v}} =  - \nabla p + \frac{1}{c} \, {\bf{j}} \times {\bf{B}} + \rho \nu {\nabla ^2}{\bf{v}} \, ,\label{e1}
\end{equation}
whereas the induction equation reads
\begin{equation}
\begin{array}{*{20}{c}}
\, \\
\, 
\end{array} 
{\partial_t}{\bf{B}} = \nabla  \times \left( {{\bf{v}} \times {\bf{B}}} \right) + {D_\eta }{\nabla ^2}{\bf{B}} \, , \label{e2}
\end{equation}
where $\bf{v}$ and $p$ are the velocity and pressure of the plasma, respectively, $\bf{B}$ is the magnetic field, ${\bf{j}} = (c/4\pi) \nabla  \times {\bf{B}}$ denotes the electric current density, and $\rho$ stands for the mass density. The kinematic viscosity can be recognized as $\nu$, while $D_\eta = \eta c^2 /4\pi$ is the magnetic diffusivity. If we further assume that the plasma is incompressible and homogeneous, then the mass density is constant in space and time, and the conservation of mass $\partial_t \rho + \nabla  \cdot (\rho {\bf{v}}) = 0$ reduces to the condition $\nabla  \cdot {\bf{v}} = 0$.

We also consider a two-dimensional dynamics with $\partial_z = 0$ for all the fields. In the presence of a strong and nearly constant guide field ${B_z}$, and if $\beta = p/\left( {B_z^2/8\pi } \right) \ll 1$, the components $v_z$ and $B_z$ decouple from ${{\bf{v}}_\bot}$ and ${{\bf{B}}_\bot}$. In this case, after the normalization 
\begin{equation}
\begin{array}{*{20}{c}}
\, \\
\, 
\end{array} 
\left( {L\nabla, \, \frac{t}{\tau_A}, \, \frac{\bf{B}}{B_0}} \right) \to \left( {\nabla, \, t, \, {\bf{B}}} \right) ,\label{e3}
\end{equation}
where $\tau_A = L/(B_0/ \sqrt {4\pi \rho }) \,$, and after introducing the stream and flux functions $\phi(x,y,t)$ and $\psi(x,y,t)$ such that the normalized in-plane velocity can be written as ${{\bf{v}}_ \bot } = {{\bf{e}}_z} \times \nabla \phi$ and the normalized in-plane magnetic field as ${{\bf{B}}_ \bot } = \nabla \psi  \times {{\bf{e}}_z}$, the equations (\ref{e1}) and (\ref{e2}) may be reduced to \cite{Strauss_1976}
\begin{equation}
\begin{array}{*{20}{c}}
\, \\
\,
\end{array} 
{\partial_t}{\omega_z} + \left[ {\phi ,{\omega_z}} \right] = \left[ {{j_z} ,\psi} \right] + \nu {\nabla^2}{\omega_z} \, , \label{e4}
\end{equation}
\begin{equation}
\begin{array}{*{20}{c}}
\, \\
\, 
\end{array} 
{\partial_t}\psi  + \left[ {\phi ,\psi } \right] =  - \eta j_z + E_0 \, . \label{e5}
\end{equation}
The canonical Poisson brackets are given as usual by $\left[ {f,g} \right] = {\partial_x}f{\partial_y}g - {\partial_y}f{\partial_x}g$. Furthermore, $\omega_z = \nabla^2_\bot \phi$, $\, j_z = - \nabla^2_\bot \psi$, and $\nabla_\bot = {{\bf e}_x} \partial_x + {{\bf e}_y} \partial_y$. Notice also that in Eq. (\ref{e5}) we have added a term representing an externally applied electric field $E_{0} = \eta j_{z}^{(0)}$, which is required to maintain the equilibrium magnetic field in the presence of a small but finite electrical resistivity $\eta$.

In accordance with the Taylor problem, we assume a slab plasma which is bounded by perfectly conducting walls at $x=\pm L_x/2$ and is periodic in the $y$ direction with period $L_y$. Furthermore, we consider a tearing-stable equilibrium given by
\begin{equation}
\begin{array}{*{20}{c}}
\, \\
\, 
\end{array} 
{\psi ^{(0)}}(x) =  - x^2/2  \, , \qquad {\phi ^{(0)}}(x) = 0 \, . \label{equil}
\end{equation}
Therefore, ${B_y}^{(0)}(x) = x$, ${j_z}^{(0)}(x) = 1$, and ${B_x}^{(0)}(x) = {v_x}^{(0)}(x) = {v_y}^{(0)}(x) = {\omega_z}^{(0)}(x) = 0$. For this set-up we have (in unnormalized units) $L=L_x$ and $B_0=\Delta B_y^{(0)}={B_y}^{(0)}(L_x/2)-{B_y}^{(0)}(-L_x/2)$.

Then, we suppose that the conducting walls are subject to a sudden displacement 
\begin{equation}
\begin{array}{*{20}{c}}
\, \\
\, 
\end{array} 
x_w =  \pm 1/2 \mp {\Xi (t)}\cos (ky) \ ,
\end{equation}
where $k=2\pi/L_y$ and $\Xi (t) = {\Xi _0}\left( {1 - {e^{ - t/\tau }} - (t/\tau ){e^{ - t/\tau }}} \right)$ for $t \geq 0$ \cite{Fitz_2004}. Since the displacement is small, i.e. ${\Xi_0} \ll 1/2$, we can adopt the following boundary conditions at the flux conserving walls
\begin{equation}
\psi ( \pm 1/2,y,t) =  - 1/8 - \Psi (t)\cos (ky) \, , \qquad   j_z( \pm 1/2,y,t) = 1 \, , \label{BC1}
\end{equation}
\begin{equation}
\begin{array}{*{20}{c}}
\, \\
\, \\
\,
\end{array} 
\phi ( \pm 1/2,y,t) =  \pm \frac{{d\Xi (t)}}{{dt}}\frac{{\sin (ky)}}{k} \, , \qquad   \omega_z( \pm 1/2,y,t) = 0 \, , \label{BC2}
\end{equation}
where $\Psi (t) = {\left. {\dfrac{{d{\psi ^{(0)}}}}{{dx}}} \right|_{x =  - {1}/2}}\Xi (t)$.

\section{Reconnection rate and plasmoid formation}

We solve Eqs. (\ref{e4}) and (\ref{e5}) numerically by splitting all the fields in the time independent equilibrium and an evolving perturbation, which is advanced in time according to a third order Adams-Bashforth algorithm. We employ a compact finite difference algorithm on a non-equispaced grid to compute the spatial operations in the $x$ direction, while a pseudospectral method is used for the periodic direction. We adopt a space discretization of 512$\times$8192 grid points, so that the minimum step size in the $x$ direction is $d{x_{\min }} = 8.04 \times {10^{ - 4}}$, whereas in the $y$ direction we have $dy = 3.07 \times {10^{ - 3}}$ ($L_y=8\pi$). Convergence studies have been conducted to ensure that the results do not change when increasing the resolution.

In the following, we consider the results of two simulations with the same boundary perturbation ${\Psi _0} = {\left. {d{\psi ^{(0)}}/dx} \right|_{x =  - 1/2}}{\Xi _0}$, but different values of the plasma resistivity and viscosity. These values ​​correspond to $\eta = 10^{ - 5}, \, \nu = 10^{ - 4},$ in one case, and $\eta = 10^{ - 8}, \, \nu = 10^{ - 7},$ in the other case. Therefore, while the Lundquist number $S= 1 / \eta$ is different, the Prandtl number $P_m = \nu / \eta$ does not change.

\begin{figure}
\begin{center}
\includegraphics[bb = 18 216 576 628, height=5.62cm]{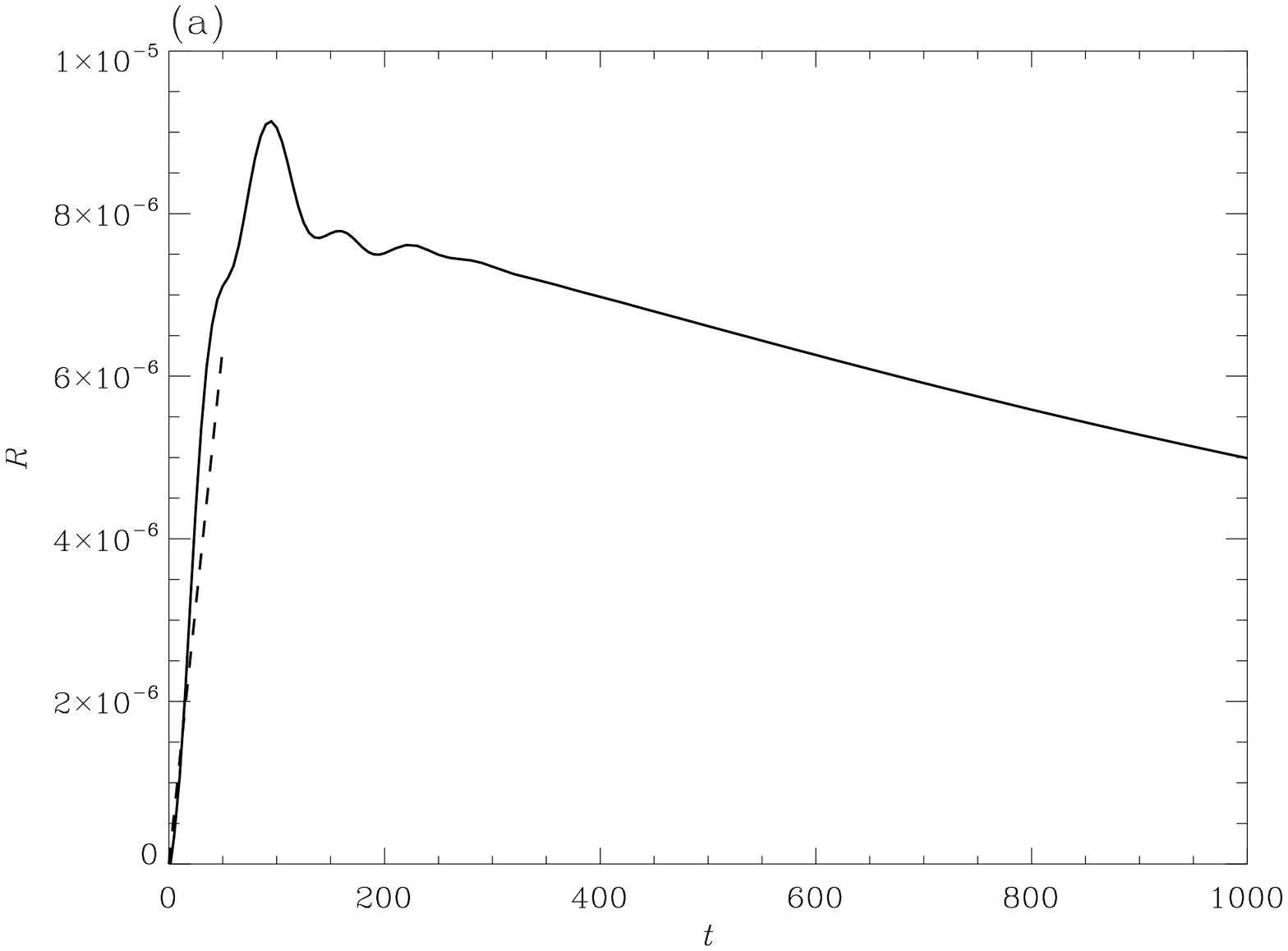}
\includegraphics[bb = 18 214 577 630, height=5.62cm]{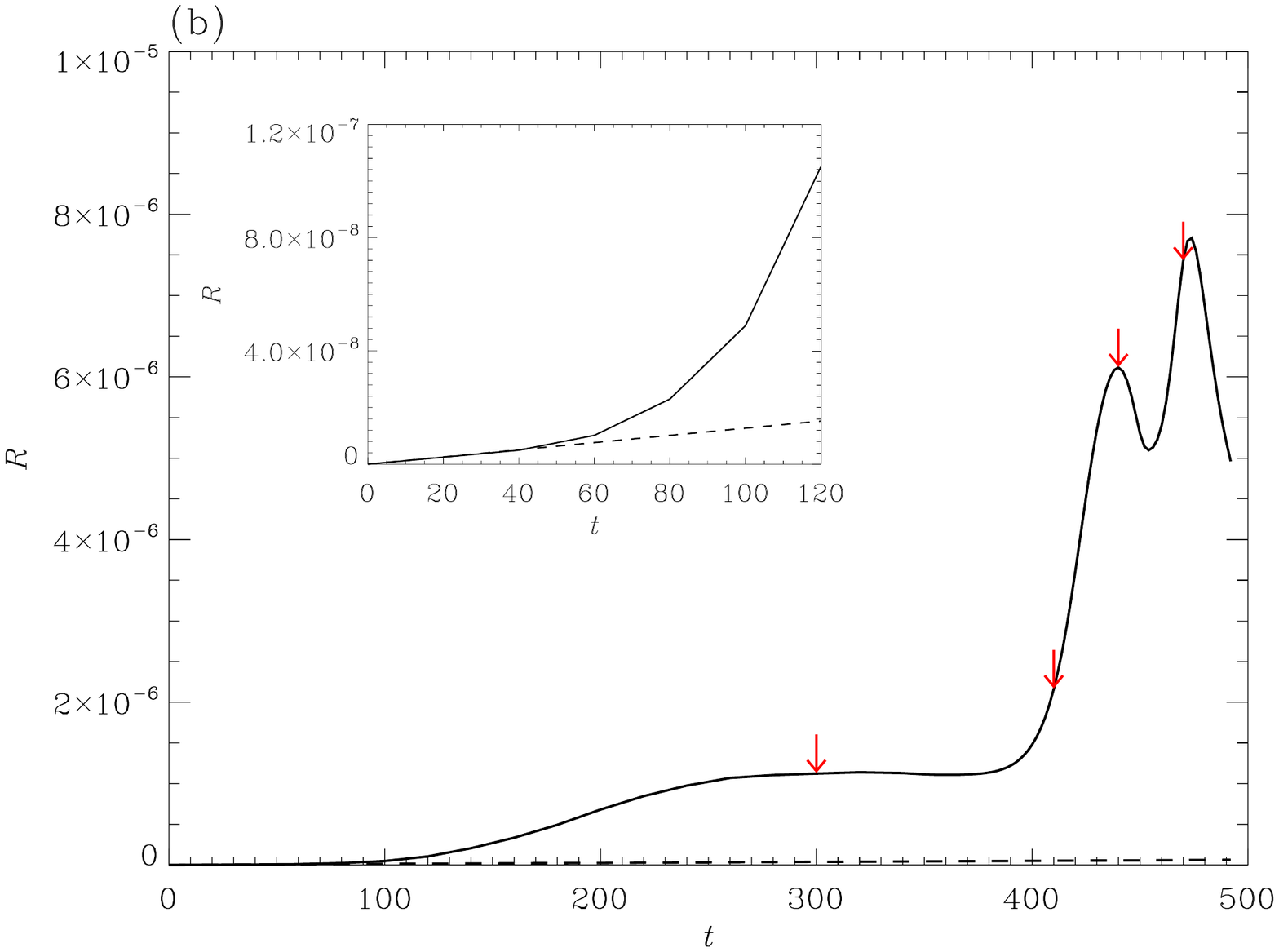}
\end{center}
\caption{Reconnection rate $R(t)$ for (a) $\eta = 10^{ - 5}$, $\nu = 10^{ - 4}$, and (b) $\eta = 10^{ - 8}$, $\nu = 10^{ - 7}$. Furthermore, $\Psi_0= 1 \times {10^{ - 2}}$, $\tau=1$, and $L_y=8\pi$. Therefore, the usual tearing stability parameter for the equilibrium (\ref{equil}) and wave number $k=2\pi/L_y$ is $\Delta ' =  - 2k/\tanh \left( {k\,{L_x}/2} \right)  \approx - 4.02$. The theoretical reconnection rates in the inertial regime are represented by dashed lines. The inset shows a zoom at early times. Red arrows correspond to the time when the current density contours are shown in Fig. \ref{fig2}.}
\label{fig1}
\end{figure}

Fig. \ref{fig1} shows clearly that the time evolution of the reconnection rate, defined as
\begin{equation}
\begin{array}{*{20}{c}}
\, \\
\, 
\end{array} 
R(t) = \frac{d}{{dt}} \bigg( \max \big( {\psi (0,y,t)} \big)  - \min \big( {\psi (0,y,t)} \big)  \bigg) \, , \label{Rec_Rate}
\end{equation}
is very different in the two simulations. As we shall see later, this is because the forced reconnection process evolves into a regime characterized by the formation of plasmoid chains when the plasma resistivity and viscosity are very small.

In both cases there is an initial inertial phase for $t \ll \nu ^{-1/3} k^{-2/3}$. Neglecting the very small initial transient, the perturbed flux function at the $X$-point is \cite{HK_1985}
%
\begin{equation}
\begin{array}{*{20}{c}}
\, \\
\, 
\end{array} 
\delta {\psi_{\rm{i}}}(t) = \frac{\eta }{\pi }\frac{{2{k^2}{\Psi _0}}}{{\sinh \left( {k\,{L_x}/2} \right)}}\,{t^2} \, . \label{Rec_Flux_I}
\end{equation}
Therefore, since in this phase the $X$ and $O$ points of the driven magnetic island are located in correspondence of the maximum and minimum of the perturbed flux function, the reconnection rate defined by Eq. (\ref{Rec_Rate}) can be evaluated as
\begin{equation}
\begin{array}{*{20}{c}}
\, \\
\, 
\end{array} 
{R_{\rm{i}}}(t) = \frac{\eta }{\pi }\frac{{8{k^2}{\Psi _0}}}{{\sinh \left( {k\,{L_x}/2} \right)}}\,t \, . \label{Rec_Rate_I}
\end{equation}
From Fig. \ref{fig1} we can see that in both simulations the initial phase evolves in accordance with the above formula represented in the plots by dashed lines. During this inertial phase, which is characterized by a nonconstant-$\psi$ behaviour, a current layer builds up along the resonant surface $x=0$.

After the inertial phase, the evolution of the system can be very different depending on the magnitude of the boundary perturbation with respect to that of the relevant plasma parameters. Fig. \ref{fig1}(a) shows the reconnection evolving as predicted by Hahm and Kulsrud \cite{HK_1985}. In this case, the inertial phase is followed by a linear constant-$\psi$ phase. For $P_m = \nu / \eta \gg 1$ this phase is known as the visco-resistive regime, which occurs for $t \gg \nu^{1/3} \eta ^{-2/3} k^{-2/3}$ \cite{Fitz_2003}. 
In our simulation the system remains in the visco-resistive regime until full reconnection is achieved. However, if the amplitude of the boundary perturbation is sufficiently large to allow the system to enter into the nonlinear regime, the constant-$\psi$ phase is followed by a Rutherford evolution \cite{Rutherford_1973,HK_1985,Fitz_2003}.

On the contrary, if the driven magnetic island is comparable or exceeds the linear layer width while it is still in the nonconstant-$\psi$ phase, the system passes into the nonlinear regime (at $t \approx 200$ for the simulation shown in Fig. \ref{fig1}(b)) according to a Sweet-Parker evolution instead of a Rutherford one \cite{WB_1992}. In this case the magnetic island is characterized by a helicity-conserving dynamics that produces an elongated current sheet \cite{Waelbroeck_1989} and leads to a reconnection rate $R \propto {\eta ^{3/4}}{\nu ^{ - 1/4}}$ for $P_m \gg 1$ \cite{Park_1984}. The threshold perturbation amplitude required to trigger the Sweet-Parker phase is \cite{Fitz_2003}
\begin{equation}
\begin{array}{*{20}{c}}
\, \\
\, 
\end{array} 
\Psi_0 \gtrsim \frac{{2k}}{{\sinh \left( {k\,{L_x}/2} \right)}}\frac{{{{(\eta \nu )}^{1/6}}}}{{{k^{1/3}}}} \, ,
\label{threshold}
\end{equation}
which tells us that in high-performance tokamaks the Sweet-Parker phase is facilitated because of the smallness of the plasma resistivity and viscosity.

\begin{figure}
\begin{center}
\includegraphics[bb = 242 224 337 734, height=21.08cm]{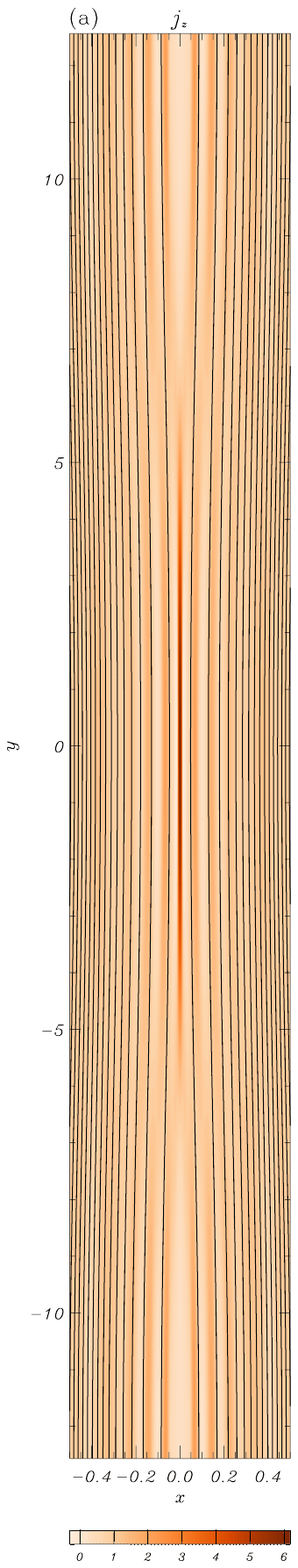}
\includegraphics[bb = 250 224 337 734, height=21.08cm]{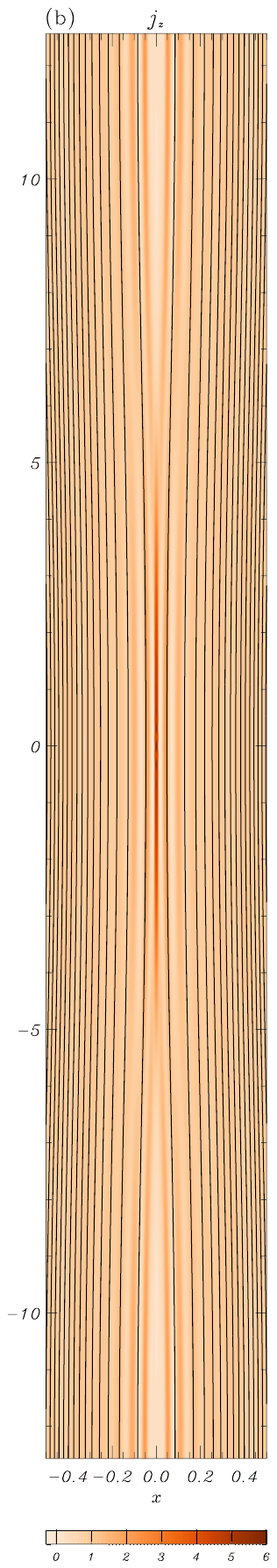}
\includegraphics[bb = 250 224 337 734, height=21.08cm]{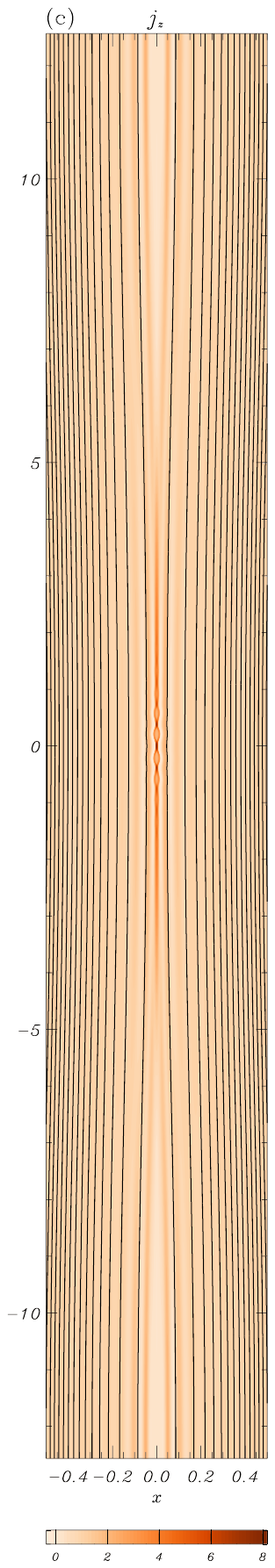}
\includegraphics[bb = 250 224 337 734, height=21.08cm]{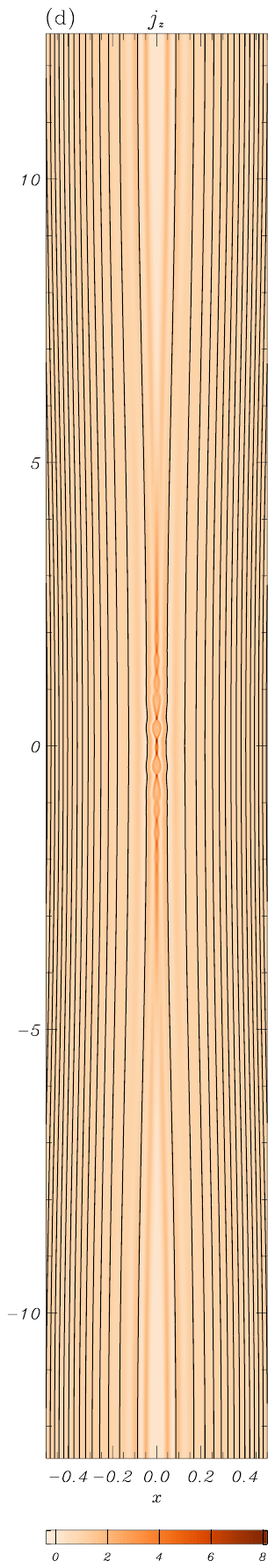}
\end{center}
\caption{From the numerical simulation shown in Fig. \ref{fig1}(b), contour plots of the out-of-plane current density $j_z$ with the in-plane component of some magnetic field lines (black lines) superimposed at (a) $t=300$, (b) $t=410$, (c) $t=440$ and (d) $t=470$.}
\label{fig2}
\end{figure}

Wang and Bhattacharjee concluded that the Sweet-Parker phase gives way to a slower Rutherford regime \cite{WB_1992}. However, this is only one possible scenario. In fact, it is important to point out that in their analysis it was implicitly assumed a stable Sweet-Parker current sheet. This is a strong assumption, which may be violated for small values of plasma resistivity and viscosity. In this case, as shown in Fig. \ref{fig2}, the thin current sheet of width $\delta_{g} \propto {\eta^{1/2}} {(\nu /\eta )^{1/4}}$ \cite{Park_1984} becomes unstable to a chain of plamoids, leading to a strong increase of the reconnection rate.

Fig. \ref{fig2}(a) shows the Sweet-Parker current sheet before the onset of the instability, which occurs at $t \approx 350$. The plasmoids closer to the center of the current sheet grow at a faster rate than those more distant from it; therefore, the four central plasmoids pass into the nonlinear phase at $t \approx 410$ while the other plasmoids are still in the linear growth phase. As shown in Fig. \ref{fig1}(b), approximately at this time there is a sudden increase of the reconnection rate. After $t \approx 440$, when many plasmoids are well into the nonlinear phase, the reconnection rate seems to fluctuate around a constant mean value, as expected on the basis of previous numerical simulations of nonlinear reconnection with plasmoids (see, e.g., \cite{BHYR_2009,Cassak_2009,HB_2010}). The growth of the plasmoids is so fast that they start to merge before being ejected in the downstream region, and the simulation is stopped at $t = 490$, when our truncated Fourier expansion can no longer resolve the large gradients in the $y$ direction that are due to the coalescence of the plasmoids in this stage of the reconnection process.

\begin{figure}
\begin{center}
\includegraphics[bb = 240 224 342 734, height=20.81cm]{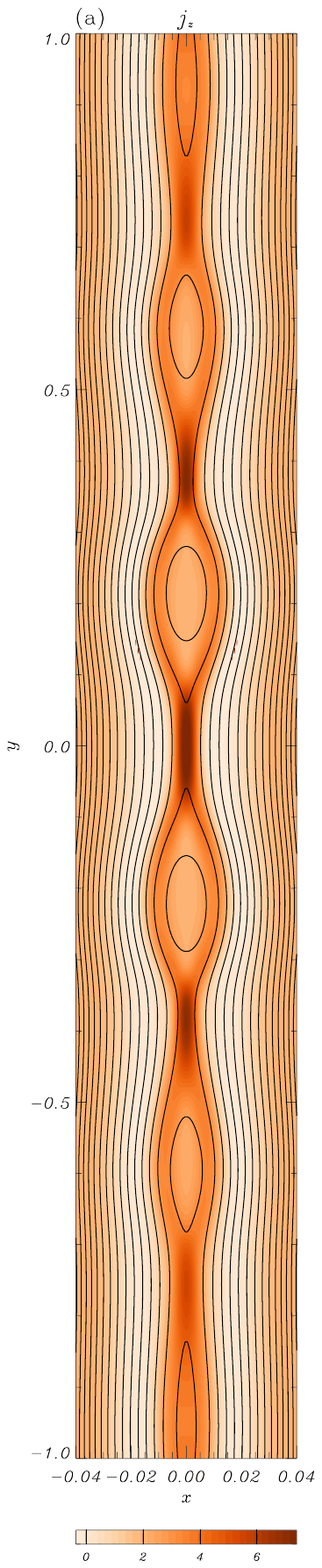}
\includegraphics[bb = 248 223 342 734, height=20.81cm]{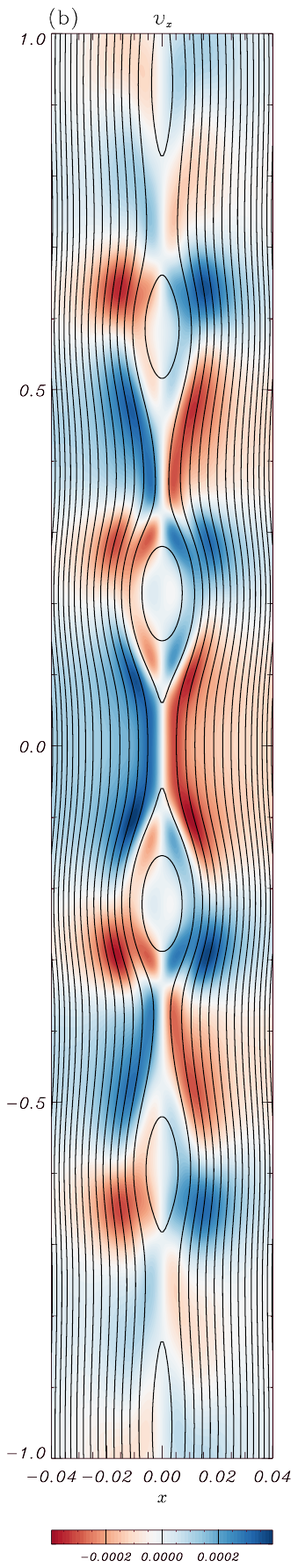}
\includegraphics[bb = 248 223 342 734, height=20.81cm]{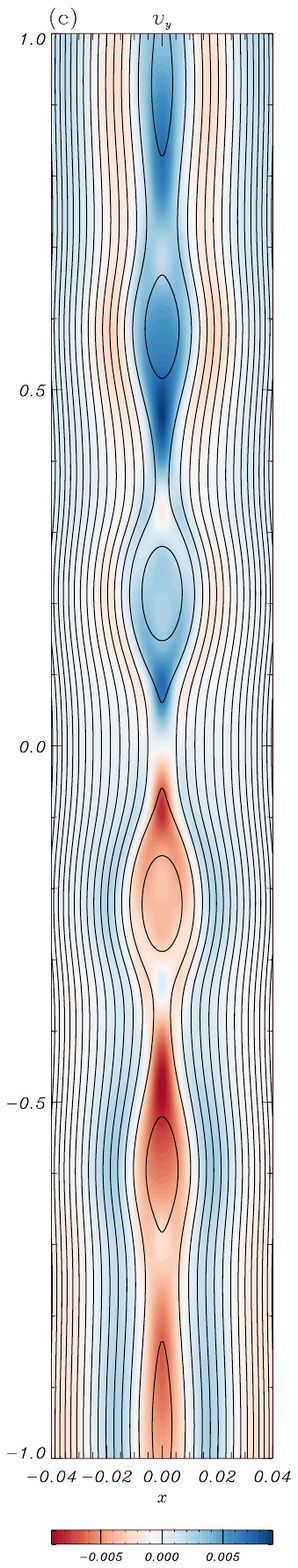}
\includegraphics[bb = 248 223 342 734, height=20.81cm]{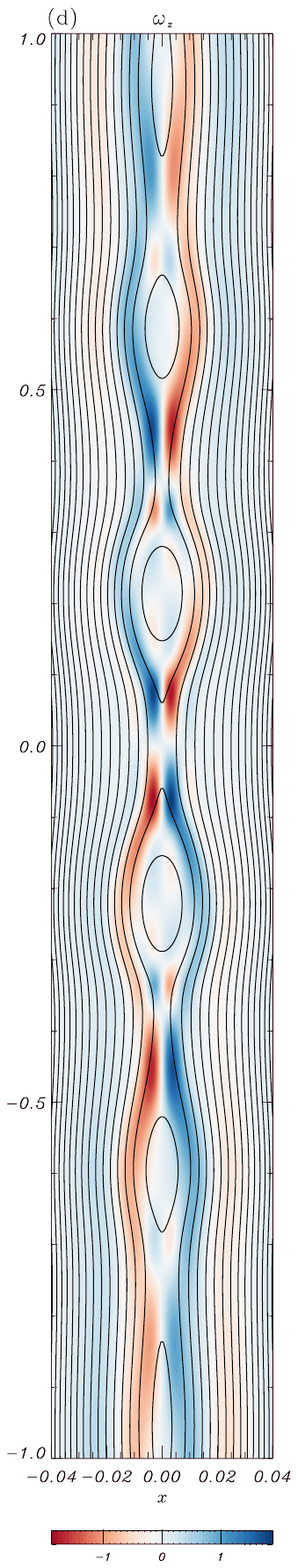}
\end{center}
\caption{From the numerical simulation shown in Fig. \ref{fig1}(b), blowup around the central plasmoids of the (a)~out-of-plane current density $j_z$, (b)~velocity $v_x$, (c)~velocity $v_y$ and (d)~vorticity $\omega_z$ at $t=440$. The in-plane component of some magnetic field lines have been superimposed (black lines).}
\label{fig3}
\end{figure}

Fig. \ref{fig3}(a) shows a zoom around the central plasmoids of $j_z$ at $t=440$. The initial Sweet-Parker current sheet is broken into smaller sheets separated by plasmoids. The sheets closer to the center are characterized by higher current density than those more distant from it. Furthermore, Fig. \ref{fig3}(b) shows that the field $v_x$ is characterized by a multipolar structure close to $x=0$, indicating that the presence of plasmoids leads to a temporary deflection of the plasma from the outflow channel. This yields a non-monotonic velocity $v_y$ at $x=0$, as shown in Fig. \ref{fig3}(c). 
It is also worth noting that, apart for the central microscopic current sheet, the other sheets are characterized by an asymmetric outflow that may lead to a kind of single wedge shape \cite{Murphy_2010}. Finally, Fig. \ref{fig3}(d) shows that the plasmoids do not develop net vorticity, differently from what happens when the reconnecting magnetic fields are not symmetric \cite{Murphy_2013}.

It is possible to obtain a rough estimation of the reconnection rate in the highly nonlinear regime by assuming a chain of plasmoids connected by various marginally stable current sheets of length $L_{c}$ and width $\delta_{c}$. Following Huang and Bhattacharjee \cite{HB_2010}, $L_{c}$ and $\delta_{c}$ may be deduced by supposing a statistical steady state in which the interplasmoid current layers follow the same Sweet-Parker scaling of the primary (global) current sheet, but with reduced length. Therefore, for $P_m \gg 1$ one can obtain
\begin{equation}
\begin{array}{*{20}{c}}
\, \\
\, 
\end{array} 
{L_{c}} \sim \eta \, \frac{{{S_{c}}}}{{{v_{A,up}}}} \sim {L_{g}} \, \frac{{{S_{c}}}}{{{S_{g}}}} \, ,
\label{Critical_lenght}
\end{equation}
and
\begin{equation}
\begin{array}{*{20}{c}}
\, \\
\, 
\end{array} 
{\delta _c} \sim \frac{{{L_c}}}{{S_c^{1/2}}}P_m^{1/4} \sim {L_g}\frac{{S_c^{1/2}}}{{{S_g}}}P_m^{1/4} \, ,
\label{Critical_width}
\end{equation}
where $S_{g}={v_{A,up}} L_{g} / \eta$ is the Lundquist number based on the global current sheet lenght $L_g$, $v_{A,up}$ is the Alfv\'en speed based on the reconnecting component of the magnetic field $B_{y,up}$, and $S_c$ is the critical Lundquist number above which the current sheet is unstable to the plasmoid instability. 
The reconnection rate may be estimated simply as the rate of change of the flux reconnected via the most central layer \cite{ULS_2010}. Hence it follows that
\begin{equation}
\begin{array}{*{20}{c}}
\, \\
\, 
\end{array} 
R \sim \eta \, \frac{{{B_{y,up}}}}{{{\delta_{c}}}} \sim \frac{{{B_{y,up}}\,{v_{A,up}}}}{{S_c^{1/2}P_m^{1/4}}} \, ,
\label{Rec_rate_plasmoids}
\end{equation}
showing that $R$ is intimately linked to the aspect ratio of the elementary current sheet of the plasmoid chain. When the plasma viscosity is negligibe, i.e. $P_m \ll 1$, numerical simulations indicate that ${S_c} \sim {10^4}$ \cite{Bisk_1986}. Thus, since in the absence of viscosity Eq. (\ref{Rec_rate_plasmoids}) has to be replaced by \cite{HB_2010} $R \sim S_c^{ - 1/2}{B_{y,up}}\,{v_{A,up}}$, in this regime the statistical steady state reconnection rate 
is independent of the microscopic plasma parameters. For $P_m \gg 1$ there are not yet numerical indications for $S_c$. However, heuristic arguments by Loureiro and coworkers suggest ${S_c} \sim {10^4}{P_m}^{1/2}$ \cite{LSU_2013}. Therefore, the reconnection rate in the visco-resistive regime can be estimated as
\begin{equation}
\begin{array}{*{20}{c}}
\, \\
\, 
\end{array} 
R  \sim {10^{ - 2}}{P_m}^{ - 1/2}{B_{y,up}} \, {v_{A,up}} \, .
\label{Rec_rate_plasmoids_2}
\end{equation}
A numerical check of the above formula would require to perform several numerical simulations at different magnetic Prandtl numbers. We leave such study for a future work. However, it is worth to compare the prediction of this formula with the reconnection rate of the highly nonlinear plasmoid regime shown in Fig. \ref{fig1}(b). In this stage of the reconnection process the reconnecting component of the magnetic field just upstream of the diffusion region is $B_{y,up} \approx  0.032$. Then, the relation (\ref{Rec_rate_plasmoids_2}) gives $R  \sim 3.24 \times {10^{-6}}$, which is reasonably close to the numerical reconnection rate. According to the scalings of Park {\it et al.} \cite{Park_1984}, in the absence of plasmoids the value of the reconnection rate for this small-resistivity small-viscosity case would be much lower than the one observed in Fig. \ref{fig1}(a); in contrast, the plasmoids increase the reconnection rate so strongly that the maximum reconnection rate is essentially the same despite the large difference in the values of plasma resistivity and viscosity.

\section{Conclusions}

In this paper we have found a new nonlinear scenario of the Taylor problem which complements those of Hahm and Kulsrud \cite{HK_1985} and Wang and Bhattacharjee \cite{WB_1992}. For very small boundary perturbations Hahm and Kulsrud \cite{HK_1985} demonstrated that the driven magnetic island evolves into the nonlinear phase through a slow Rutherford regime, whereas Wang and Bhattacharjee \cite{WB_1992} showed that larger boundary perturbations lead to a nonlinear evolution characterized by a Sweet-Parker phase that subsequently gives way to the Rutherford regime. Here instead we have shown that larger boundary perturbations may give rise to a Sweet-Parker phase with a current sheet that is unstable to the plasmoid instability \cite{Lou_2007} if very small values of plasma resistivity and viscosity are considered. As a consequence, the advanced nonlinear phase is characterized by the development of a plasmoid chain that is responsible for a substantial speed up of the reconnection process. When the plasmoid chain is fully developed, the reconnection rate in the large magnetic Prandtl number regime is estimated to be $R \sim {10^{-2}}{P_m}^{-1/2}{B_{y,up}} \, {v_{A,up}}$. A  detailed scaling study of the reconnection rate in the visco-resistive regime is the subject of ongoing work, as well as the determination of an expression for the threshold perturbation required to trigger the plasmoid phase.

\ack
The authors would like to acknowledge fruitful conversations with Dario Borgogno, Richard Fitzpatrick, Enzo Lazzaro and Fulvio Militello. One of us (L.C.) is grateful for the hospitality of the Institute for Fusion Studies at the University of Texas at Austin, where part of this work was done.
This work was carried out under the Contract of Association Euratom-ENEA and was also supported by the U.S. Department of Energy under Contract No. DE-FG02-04ER-54742.

\end{document}